\documentclass[aps,prl,amsmath,amssymb,twocolumn,floatfix,10pt,showpacs]{revtex4-1}
\usepackage{graphicx}
\usepackage{hyperref}
\usepackage{color}

\newcommand{\ve}{\mathbf}

\begin{document}

\title{Anisotropic super-spin at the end of a carbon nanotube}
\author{Manuel J. Schmidt}
\affiliation{Department of Physics, University of Basel, Klingelbergstrasse 82, 4056 Basel, Switzerland}
\date{\today}


\begin{abstract}
Interaction-induced magnetism at the ends of carbon nanotubes is studied theoretically, with a special focus on magnetic anisotropies. Spin-orbit coupling, generally weak in ordinary graphene, is strongly enhanced in nanotubes. In combination with Coulomb interactions, this enhanced spin-orbit coupling gives rise to a super-spin at the ends of carbon nanotubes with an XY anisotropy on the order of 10 mK. Furthermore, it is shown that this anisotropy can be enhanced by more than one order of magnitude via a partial suppression of the super-spin.
\end{abstract}

\maketitle

{\it Introduction.} Carbon-based material systems are promising candidates for applications in future (quantum) information technologies and attract much attention. One reason for their great potential is that electrons on carbon honeycomb lattices behave unusually in many respects. For instance, instead of being quasiparticles resembling free electrons, the low-energy excitations in graphene and in some carbon nanotubes (CNTs) are described by the Dirac theory \cite{graphene_rmp_2009}. An important consequence of this Dirac nature is the efficient suppression of Coulomb interaction effects, even though, due to the extreme electronic confinement to only one atomic layer, the energy scale corresponding to this interaction is rather large on two-dimensional carbon honeycomb lattices \cite{graphene_interactions_rmp_2009}. As a consequence, non-interacting theories of graphene or CNTs are very successful as long as the honeycomb lattice is intact; in order to drive a phase transition in bulk graphene, even larger interaction strengths are required \cite{meng_spin_liquid_2010}.

Where the honeycomb lattice is disturbed, however, the Coulomb interaction has a chance to manifest itself. Examples of this mechanism are the vacancy-induced magnetism in graphite \cite{esquinazi_proton_irradiaion_graphite_magnetism_2003} or edge magnetism, which appears at zigzag edges of graphene flakes \cite{son_edge_magnetism_2006}. In these examples, disturbances of the honeycomb lattice (vacancies and edges, respectively) allow interactions to drive a magnetic transition. In the present work, a similar magnetic transition at the ends of CNTs \cite{kim_cnt_magnetism_2003} is studied: the Coulomb interaction aligns the spins of all electrons in certain states, localized at the CNT end, and thus gives rise to a super-spin, composed of many individual electron spins. This CNT end magnetism is the pendant of graphene's edge magnetism \cite{son_edge_magnetism_2006}. 

Usually, the spin-orbit interaction (SOI) is assumed to be negligible for edge magnetism. This is because the SOI energy scale in graphene (28 $\mu$eV \cite{gmitra_soi_2009}) is much smaller than the electron-electron interaction ($\sim$ eV \cite{wehling_strength_U_2011}). Moreover, the rigidity of the super-spin suppresses SOI effects even further so that their effective energy scale is in the $\mu$K regime, and this renders the SOI experimentally irrelevant. However, it is known that the SOI can be enhanced: at graphene/graphane interfaces, for instance, spin-orbit effects have been shown to be amplified by two orders of magnitude \cite{schmidt_gg_soi_2010}. Also the surface curvature of CNTs gives rise to an enhanced SOI \cite{ando_soi_cnt_2000}. 

In this Letter, the spin-orbit anisotropy of the super-spin at CNT ends is calculated on the basis of a microscopic model. It is shown that the curvature-induced enhancement of the SOI lifts the anisotropy up to an experimentally accessible regime. Moreover, by a combination with the mechanism for tuning the strength of edge magnetism \cite{schmidt_tem_2010}, which translates to a tuning of the size of the super-spin at CNT ends, the anisotropy can be further increased.

\begin{figure}[!ht]
\centering
\includegraphics[width=\columnwidth]{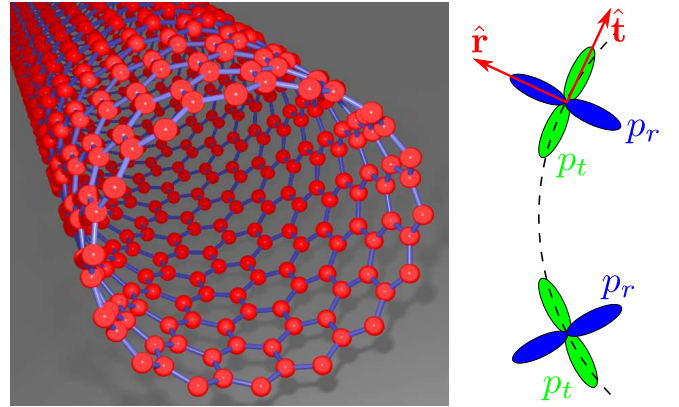}
\caption{(Color online) {\it Left:} the end of a 15-CNT, i.e., a CNT with $N_c=15$ unit cells in circumferential direction. {\it Right:} the local coordinate system $\{\hat{\ve r},\hat{\ve t},\hat{\ve z}\}$ ($\hat{\ve z}$ points into the paper sheet) and the orientation of the $p$ orbitals $p_r$ and $p_t$ at the surface of the CNT (dashed line).}
\label{fig_cnt}
\end{figure}

{\it Model.} This study starts from an interacting model for the $\pi$ and $\sigma$ orbitals of the CNT's carbon atoms. The Hamiltonian consists of three parts
\begin{equation}
H = H^{\rm hop} + H^{\rm SO} + H^{\rm int}.
\end{equation}
The hopping between different orbitals of nearest-neighbor atoms is described by $H^{\rm hop} = t^{ij}_{\mu\mu'} c_{i\mu\sigma}^\dagger c_{j\mu'\sigma} + \epsilon_s c^\dagger_{is\sigma}c_{is\sigma}$, where the fermionic operator $c_{i\mu\sigma}$ annihilates an electron with spin $\sigma$ in orbital $\mu$ at site $i$ of a honeycomb lattice rolled up to a CNT. Here and henceforth, repeated indices are to be summed over. $t_{\mu\mu'}^{ij}$ is the amplitude for an electron hopping from orbital $\mu'$ at site $j$ to orbital $\mu$ at site $i$. All second shell orbitals $\mu=s,p_r,p_t,p_z$ of the carbon atoms are taken into account. The $p$ orbitals are defined in a local coordinate system $\{\hat{\ve r},\hat{\ve t},\hat{\ve z}\}$ (see Fig. \ref{fig_cnt}). $\epsilon_s=-8.9$ eV is the energy of the $s$ orbital. The CNT curvature is encoded in $t_{\mu\mu'}^{ij}$ in a highly nontrivial manner, described in detail in Refs. \onlinecite{schmidt_gg_soi_2010,klinovaja_cnt_long_2011}. The SOI reads $H^{\rm SO} = i\Delta_{\rm SO} \epsilon^{\mu\nu\eta} c^\dagger_{i\mu\sigma} s^{\nu}_{\sigma\sigma'} c_{i\eta\sigma'}$ \cite{schmidt_gg_soi_2010}, with $\Delta_{\rm SO}=6$ meV \cite{serrano_soi_strength_abinitio_2000}, $\epsilon^{\mu\nu\eta}$ is the Levi-Civita tensor and $s^\mu$ are the spin Pauli matrices (with eigenvalues $\pm1$), defined in the local coordinate system $\{\hat{\ve r},\hat{\ve t},\hat{\ve z}\}$. The contributions from the $d$ orbitals to the SOI are negligible because of the curvature-induced $\pi\sigma$ hybridization \cite{gmitra_soi_2009}. The Coulomb interaction is modeled by a Hubbard Hamiltonian for the $\pi$ orbitals only, i.e., $H^{\rm int}=U c^\dagger_{ip_r\uparrow}c_{ip_r\uparrow}c^\dagger_{ip_r\downarrow}c_{ip_r\downarrow}$, with $U\simeq 9.3$ eV \cite{wehling_strength_U_2011}. The Coulomb interaction for the $\sigma$ electrons is assumed to be included on an effective level in the hopping parameters.

{\it Edge state model.} The CNT considered in this work is half-infinite, i.e., it has only one end at which the lattice is zigzag-terminated (see Fig. \ref{fig_cnt}). Such a termination supports edge states \cite{fujita_edge_states_1996}. For describing the magnetic state at this CNT end, it is sufficient to work in a restricted Fock space, spanned by these edge states only \cite{schmidt_tem_2010}. Therefore, fermionic edge state operators $e_{k\sigma}=[\psi^e_k(i,\mu)]^* c_{i\mu\sigma}$ are introduced, where $\psi^{e}_k(i,\mu)$ is the edge state wave function with momentum $k$ in circumferential direction (along $\hat{\ve t}$). $\psi^{e}_k(i,\mu)$ is determined by the numerical diagonalization of the hopping Hamiltonian $H^{\rm hop}$. It is derived mainly from the $\pi$ orbital $p_r$. However, due to curvature-induced $\pi\sigma$ hybridization, it has also contributions from the $\sigma$ bands; these are most important for the SOI. Simple approximations for $\psi^{e}_k(i,\mu)$ taking into account only the $\pi$ band can be found in Refs. \onlinecite{schmidt_gg_soi_2010,schmidt_tem_2010,luitz_ed_2011}. These approximations are sufficient for calculating the projection of the Hubbard Hamiltonian and of the kinetic energy of the edge states (see below), but they are insufficient for the SOI.

Essentially, there are three terms describing the physics of the edge states. The kinetic energy of the edge states \cite{schmidt_tem_2010} is described by
\begin{equation}
H_e^{\rm kin} = - \Delta \sideset{}{'}\sum_{k\sigma} \mathcal N_k^2 e^\dagger_{k\sigma} e_{k\sigma},
\end{equation}
with $\mathcal N_k=\sqrt{2\cos(k-\pi)-1}$. The $k$ sum is restricted to $\frac{2\pi}3 < k < \frac{4\pi}3$ since the edge states exist only for this range of momenta in circumferential direction \footnote{The Dirac points $\frac{2\pi}3,\frac{4\pi}3$ are excluded explicitly in this work by considering CNT circumferences $N_c$ which are not multiples of 3.}. $\Delta$ is a phenomenological parameter which accounts for a variety of electron-hole symmetry (EHS) breaking terms in the Hamiltonian, such as next-nearest neighbor hopping or local perturbations at the edge \cite{schmidt_tem_2010,luitz_ed_2011}. With EHS, the edge states have exactly zero energy. Breaking EHS gives rise to a non-zero bandwidth for the edge states; interestingly, all these EHS breakings can be subsumed in one parameter $\Delta$. Note that $\Delta$ is experimentally tunable in special geometries \cite{schmidt_tem_2010}.

The projection of the honeycomb lattice Hubbard interaction $H^{\rm int}$ onto the edge state subspace reads \cite{schmidt_tem_2010}
\begin{equation}
H^{\rm int}_e = \frac{U}{N_c}  \sideset{}{'}\sum_{k,k',q} \Gamma(k,k',q): e^\dagger_{k+q\uparrow} e_{k\uparrow} e^\dagger_{k'-q\downarrow} e_{k'\downarrow}:,
\end{equation}
where $U\simeq 9.3$ eV \cite{wehling_strength_U_2011} is the Hubbard coupling constant on the 2D honeycomb lattice, $N_c$ is the number of unit cells in circumferential direction, and $\Gamma(k,k',q) = \mathcal N_{k+q}\mathcal N_k\mathcal N_{k'-q}\mathcal N_{k'}/(1-u_{k+q}^*u_k u_{k'-q}^*u_{k'})$, with $u_k=1+e^{ik}$. Again, the $k$ sum is restricted such that the momenta of all edge state operators are in the interval $]\frac{2\pi}3,\frac{4\pi}3[$.

The interplay of $H_e^{\rm kin}$ and $H_e^{\rm int}$ has been extensively discussed in view of the tunable ferromagnetism at long graphene zigzag edges \cite{schmidt_tem_2010,luitz_ed_2011}. The present work rather focuses on finite zigzag edges with periodic boundary conditions (the graphene edge is rolled up to a CNT end). Thus, $k$ is discrete with a spacing $\Delta k=2\pi/N_c$. For $\Delta\ll U$, a super-spin with a certain size $S$ is found at the CNT end and, because $H_e^{\rm kin}$ and $H_e^{\rm int}$ are SU(2) invariant, this super-spin obeys the full rotation symmetry, that is, the ground state is $(2S+1)$-fold degenerate.

The effective SOI is derived by projecting $H^{\rm SO}$ onto the Hilbert space spanned by the edge states. Since $H^{\rm SO}$ is a single particle operator this projection is easily performed
\begin{equation}
H_e^{\rm SO} = \sideset{}{'}\sum_{k,k',\sigma,\sigma'} \left<\psi^e_{k},\sigma \right| H^{\rm SO} \left|\psi^e_{k'},\sigma'\right> e^\dagger_{k\sigma} e_{k'\sigma'},
\end{equation}
where $\left|\psi^e_k,\sigma\right>$ is the wave function of an edge state with momentum $k$ and spin $\sigma$. $H^{\rm SO}$ can be separated into a part $H^{\rm SO,z}$ proportional to $s^z$, and a part $H^{\rm SO,xy}$ proportional to $s^x$ or $s^y$. Because $H^{\rm SO}$ depends on the locally defined $s^{r,t,z}$, the terms $H^{\rm SO,xy}$ have a trigonometric dependence on the azimuthal angle $\varphi$, while $H^{\rm SO,z}$ is independent of $\varphi$ \cite{klinovaja_helical_modes_2011}. Thus, the projected $H_e^{\rm SO,z}$ is diagonal in $k$, while $H_e^{\rm SO,xy}$ is not.

\begin{figure}[!ht]
\centering
\includegraphics[width=210pt]{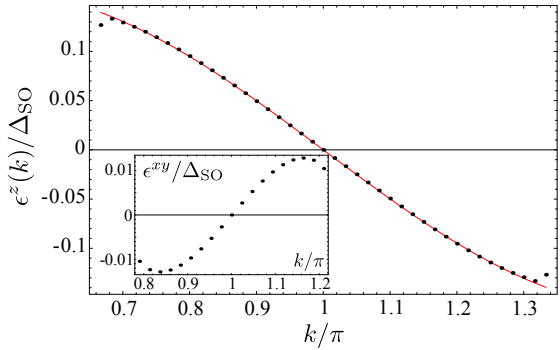}
\caption{(Color online) Numerical results (black dots) and fit $\delta_z\sin(k)$ (red line) for the effective $z$ SOI of edge states. In this figure, $N_c=10$. The deviation of the numerical $\epsilon^z$ from the sine form at $k=\frac{2\pi}3,\frac{4\pi}3$ is due to finite size effects in the numerical calculation. The inset shows the function $\epsilon^{xy}(k)$ for $N_c=10$.}
\label{fig_z_soi_fit}
\end{figure}

The most important part of the effective SOI for the super-spin anisotropy is the $k$-diagonal $H^{\rm SO,z}_e$. Due to time-reversal invariance, $\sigma\epsilon^z(k)\equiv \left<\psi^e_{k},\sigma \right| H^{\rm SO} \left|\psi^e_{k},\sigma\right>$ must be odd around $k=0$ and $k=\pi$, so that $\epsilon^z(k)$ can be expanded in $\sin(m k)$, with integer $m$. It turns out to be sufficient to keep only the leading order, i.e., $\epsilon^z(k) = \delta_z\sin(k)$. The parameter $\delta_z$ is determined by fitting $\epsilon^z(k)$ to the numerically evaluated $\left<\psi^e_{k},\uparrow \right| H^{\rm SO} \left|\psi^e_{k},\uparrow\right>$. As can be seen from Fig. 
\ref{fig_z_soi_fit}, the sine form for $\epsilon^z(k)$ is an excellent approximation so that one may write 
\begin{equation}
H^{\rm SO,z}_e = \delta_z \sideset{}{'}\sum_{k,\sigma} \sin(k) \sigma e^\dagger_{k\sigma} e_{k\sigma}.
\end{equation}
The effective spin-orbit coupling constant $\delta_z$ is fitted to the numerical results for $N_c=5,...,60$ in order to extract the $N_c$ dependence of $\delta_z$
\begin{equation}
\delta_z/\Delta_{\rm SO} \simeq \frac{1.6}{N_c} + \frac{1.1}{N_c^3}.\label{dz_fit}
\end{equation}
The relative deviation of Eq. (\ref{dz_fit}) from the numerical data is less than $10^{-4}$.

The functional form of the Hamiltonian $H_e^{\rm SO,xy}$ is less obvious. First of all, due to the trigonometric dependencies on the azimuthal angle, $H_e^{\rm SO,xy}$ is not diagonal in $k$ but couples momenta differing by $\Delta k=2\pi/N_c$. In general, one may write
\begin{equation}
H^{\rm SO,xy}_e = \sideset{}{'}\sum_k \epsilon^{xy}(k) \ve e^\dagger_{k+\Delta k/2} \cdot (s^x-is^y)\cdot \ve e_{k-\Delta k/2}+ {\rm H.c.}.
\end{equation}
$\epsilon^{xy}(k)={\rm Re}[\langle\psi^e_{k+\Delta k/2},\uparrow|H^{\rm SO}|\psi^e_{k-\Delta k/2},\downarrow \rangle]$ is again determined numerically and $\ve e^\dagger_k = (e^\dagger_{k\uparrow} , e^\dagger_{k\downarrow})$. The result is shown in the inset of Fig. \ref{fig_z_soi_fit}. Note that $\epsilon^{xy}(k)$ is significantly smaller than $\epsilon^{z}(k)$. Furthermore, since the super-spin which is generated by the Coulomb interaction has zero momentum in circumferential direction, the $k$ non-diagonal $H^{\rm SO,xy}_e$ is only of minor importance and will be neglected in the remainder of this work.

{\it Anisotropy of the super-spin.} Usually, the kinetic energy of the edge states, quantified by $\Delta$, is small compared to the interaction energy scale $U$, so that the spins of all $N\simeq N_c/3$ electrons occupying the edge states are aligned (half-filling is assumed) and form a super-spin of length $S=N/2$ \cite{luitz_ed_2011}. In the absence of SOI, this super-spin is rotationally invariant, i.e., the ground state is $(2S+1)$-fold degenerate. For the CNTs considered here, this degenerate subspace is protected by excitation energies of a few hundred meV, so that the magnetic properties of a CNT end are expected to be well described by an effective super-spin model living in this $(2S+1)$-dimensional subspace. All operators in this super-spin model are given in terms of the $(2S+1)$-dimensional representation of the angular momentum operators $\tilde S^\nu$, with $\nu=x,y,z$, for which $\sum_\nu (\tilde S^\nu)^2=S(S+1)$.

\begin{figure}[!ht]
\centering
\includegraphics[width=200pt]{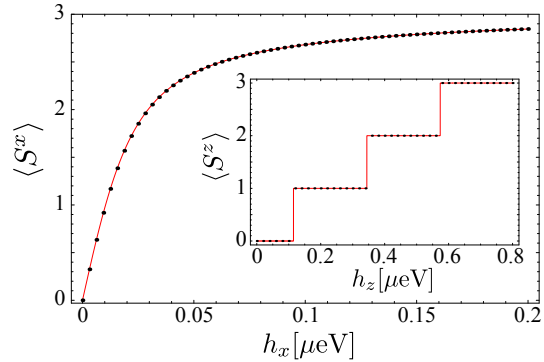}
\caption{(Color online) Spin response of the super-spin state at the end of a $N_c=17$ CNT. The dots are the results of the exact diagonalization of the interacting edge state Hamiltonian. The red line is the spin response calculated from the super-spin model Eq. (\ref{super_spin_model}).}
\label{fig_spin_response}
\end{figure}

The effective super-spin Hamiltonian $H_{\rm ss}$ is determined by requiring equal spin responses of the super-spin model $H_{\rm ss}$ and the edge state model $H_e = H_e^{\rm kin} + H_{e}^{\rm SO,z} + H_e^{\rm int}$ to a Zeeman field $H^Z= -\sum_\mu h_\mu S^\mu$, i.e., $\langle\tilde S^\nu\rangle_{\rm ss} = \langle S^\nu\rangle_e$. Here, $\langle\cdot\rangle_{\rm ss}$ denotes the ground state expectation value with respect to $H_{\rm ss}+H^Z$ and $\langle\cdot\rangle_{e}$ the ground state expectation value with respect to $H_e+H^Z$. In the edge state model, the spin operators entering the Zeeman Hamiltonian are given by $S^\nu= \frac12 \sideset{}{'}\sum_k \ve e^\dagger_{k}\cdot s^\nu \cdot \ve e_k$.

\begin{table}[!ht]
\begin{center}
\begin{tabular}{|c|c|c|c|c|c|c|c|c|c|c|}\hline
$N_c$           & 5    & 7    & 8    & 10   & 11   & 17   & 19   & 23   & 29 \\ \hline
$S$             &  1   & 1    &  3/2 & 3/2  & 2    & 3    & 3    & 4    & 5 \\
$D[\mu\rm eV]$  &  3.2 & 0.82 & 0.90 & 0.31 & 0.38 & 0.11 & 0.05 & 0.05 & 0.03 \\
$D^*[\mu\rm eV]$&      &      &      &      &      & 3.1  & 1.52 & 0.47 & 1.35 \\
$R[\rm nm]$     & 0.19 & 0.27 & 0.31 & 0.38 & 0.42 & 0.65 & 0.73 & 0.88 & 1.11\\ \hline
\end{tabular}
\end{center}
\caption{Maximal super-spin $S$, bare anisotropy constant $D$, enhanced anisotropy constant $D^*$, and radius $R$ of various zigzag CNTs with different $N_c$.}
\label{tab_d}
\end{table}

Following Ref. \onlinecite{luitz_ed_2011}, the edge state model is diagonalized exactly in order to calculate the spin response of the ground state. In the absence of SOI and Zeeman fields, the ground state is $(2S+1)$-fold degenerate ($S^z = -S,...,S$). For $\Delta_{\rm SO}\neq 0$, this degeneracy is slightly lifted. As the SOI proportional to $S^{x,y}$ is neglected, $S^z$ is a a good quantum number also in the presence of SOI, and the state(s) with the smallest $|S^z|$ forms the new ground state. For integer (half-integer) $S$, the ground state consists of $S^z=0$ ($S^z=\pm\frac12$) and is non-degenerate (two-fold degenerate). Thus, the SOI-induced anisotropy is of XY type, i.e., 
\begin{equation}
H_{\rm ss} = D (\tilde S^z)^2 \label{super_spin_model},
\end{equation}
with positive $D$, determined numerically. This is one of the main results of this work.

For a 17-CNT one finds $D=0.11\,\mu$eV by fitting $D$ to the position of the first step of the $z$-spin response. Fig. \ref{fig_spin_response} shows that the effective super-spin model describes the exact spin response extremely well. For instance, the relative differences of the positions of the steps in the $S^z$ response is typically $10^{-6}$. In the limit $\Delta_{\rm SO}\ll U$, the dependence of $D$ on the SOI and Coulomb interaction strengths is $D\sim \Delta_{\rm SO}^2/U$. Table \ref{tab_d} presents an overview of anisotropy constants for different CNTs.

\begin{figure}[!ht]
\centering
\includegraphics[width=\columnwidth]{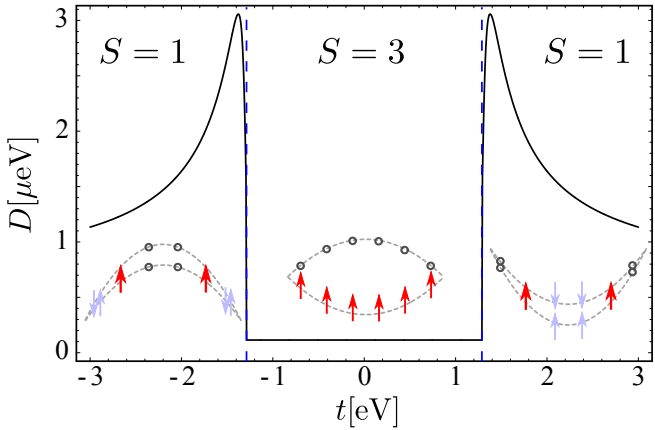}
\caption{(Color online) Increased anisotropy for the partially suppressed super-spin. For small $|\Delta|$ the size of the super-spin is $S=3$ and for larger $|\Delta|$, $S=1$. The dashed vertical lines mark the transition between these regimes. The insets show schematically the configurations of the edge state electrons: arrows correspond to occupied electrons with a certain spin -- light arrows are paired spins and bold arrows are unpaired spins; gray circles are empty states. The dashed parabolas symbolize the mean-field energies of the edge states \cite{schmidt_tem_2010}.}
\label{fig_enhanced_anisotropy}
\end{figure}

{\it Enhanced anisotropy.} The configuration in which all spins are aligned (for $|\Delta|\ll U$) corresponds to the saturated edge magnetism (SEM) regime \cite{luitz_ed_2011}. In this phase, the system is very rigid, i.e., the energy gap to the first excitation is large. This rigidity suppresses SOI effects. The weak edge magnetism (WEM) phase (for $|\Delta| \sim U$) is less rigid and has smaller excitation gaps. It is thus expected that $D$ is enhanced in the WEM phase. In the limit of infinitely long graphene edges, as it has been discussed in Refs. \onlinecite{schmidt_tem_2010,luitz_ed_2011}, it is possible to reduce the edge magnetization continuously. A CNT end, however, corresponds to a short edge and the size of the super-spin can be reduced only in steps of $\Delta S=2$. Thus, the smallest CNT for which the super-spin can be reduced to $S\geq1$ is the 17-CNT. The anisotropy enhancement will be discussed on the basis of this CNT, but Tab. \ref{tab_d} shows the enhanced anisotropies also for other CNTs with even electron numbers. The case of odd electron numbers is more complicated and beyond the scope of this work.

The dependence of $D$ on $\Delta$ is shown in Fig. \ref{fig_enhanced_anisotropy}. For small $|\Delta|$, $D$ does not depend on $\Delta$. Near the critical kinetic energy strength $|\Delta|\simeq 1.3$ eV, at which the reduction of the length of the super spin from $S=3$ to $S=1$ takes place, $D$ increases by more than an order of magnitude. The maximal $D$ is $D^*\simeq 3.1\,\mu$eV. Compared to the anisotropy of the super-spin with the maximal size (SEM regime), this is an enhancement by a factor of 27.

{\it Conclusion.} The SOI-induced anisotropy of the super-spin at CNT ends has been calculated and it was shown that this anisotropy is of XY-type. Two mechanisms have been discussed which enhance this anisotropy. One is based on the well-known curvature-induced SOI enhancement \cite{ando_soi_cnt_2000} and the other is based on the tunability of edge magnetism \cite{schmidt_tem_2010}. The first effect lifts the anisotropy of small CNTs to experimentally accessible regimes. The second effect can be used to increase the anisotropy also in larger tubes. These effects are independent. Thus, it is expected that the latter effect is also relevant in other SOI-enhanced systems supporting magnetic edge states, in particular at graphene/graphane interfaces \cite{schmidt_gg_soi_2010,schmidt_tem_2010}.

This work was financially supported by the Swiss NSF and the NCCR Nanoscience (Basel).

\vspace{-0.2cm}

\bibliography{ms}

\begin{thebibliography}{17}%
\makeatletter
\providecommand \@ifxundefined [1]{%
 \@ifx{#1\undefined}
}%
\providecommand \@ifnum [1]{%
 \ifnum #1\expandafter \@firstoftwo
 \else \expandafter \@secondoftwo
 \fi
}%
\providecommand \@ifx [1]{%
 \ifx #1\expandafter \@firstoftwo
 \else \expandafter \@secondoftwo
 \fi
}%
\providecommand \natexlab [1]{#1}%
\providecommand \enquote  [1]{``#1''}%
\providecommand \bibnamefont  [1]{#1}%
\providecommand \bibfnamefont [1]{#1}%
\providecommand \citenamefont [1]{#1}%
\providecommand \href@noop [0]{\@secondoftwo}%
\providecommand \href [0]{\begingroup \@sanitize@url \@href}%
\providecommand \@href[1]{\@@startlink{#1}\@@href}%
\providecommand \@@href[1]{\endgroup#1\@@endlink}%
\providecommand \@sanitize@url [0]{\catcode `\\12\catcode `\$12\catcode
  `\&12\catcode `\#12\catcode `\^12\catcode `\_12\catcode `\%12\relax}%
\providecommand \@@startlink[1]{}%
\providecommand \@@endlink[0]{}%
\providecommand \url  [0]{\begingroup\@sanitize@url \@url }%
\providecommand \@url [1]{\endgroup\@href {#1}{\urlprefix }}%
\providecommand \urlprefix  [0]{URL }%
\providecommand \Eprint [0]{\href }%
\providecommand \doibase [0]{http://dx.doi.org/}%
\providecommand \selectlanguage [0]{\@gobble}%
\providecommand \bibinfo  [0]{\@secondoftwo}%
\providecommand \bibfield  [0]{\@secondoftwo}%
\providecommand \translation [1]{[#1]}%
\providecommand \BibitemOpen [0]{}%
\providecommand \bibitemStop [0]{}%
\providecommand \bibitemNoStop [0]{.\EOS\space}%
\providecommand \EOS [0]{\spacefactor3000\relax}%
\providecommand \BibitemShut  [1]{\csname bibitem#1\endcsname}%
\let\auto@bib@innerbib\@empty
\bibitem [{\citenamefont {Castro~Neto}\ \emph {et~al.}(2009)\citenamefont
  {Castro~Neto}, \citenamefont {Guinea}, \citenamefont {Peres}, \citenamefont
  {Novoselov},\ and\ \citenamefont {Geim}}]{graphene_rmp_2009}%
  \BibitemOpen
  \bibfield  {author} {\bibinfo {author} {\bibfnamefont {A.~H.}\ \bibnamefont
  {Castro~Neto}}, \bibinfo {author} {\bibfnamefont {F.}~\bibnamefont {Guinea}},
  \bibinfo {author} {\bibfnamefont {N.~M.~R.}\ \bibnamefont {Peres}}, \bibinfo
  {author} {\bibfnamefont {K.~S.}\ \bibnamefont {Novoselov}}, \ and\ \bibinfo
  {author} {\bibfnamefont {A.~K.}\ \bibnamefont {Geim}},\ }\href@noop {}
  {\bibfield  {journal} {\bibinfo  {journal} {Rev. Mod. Phys.}\ }\textbf
  {\bibinfo {volume} {81}},\ \bibinfo {pages} {109} (\bibinfo {year}
  {2009})}\BibitemShut {NoStop}%
\bibitem [{\citenamefont {Kotov}\ \emph {et~al.}(2010)\citenamefont {Kotov},
  \citenamefont {Uchoa}, \citenamefont {Pereira}, \citenamefont {Castro-Neto},\
  and\ \citenamefont {Guinea}}]{graphene_interactions_rmp_2009}%
  \BibitemOpen
  \bibfield  {author} {\bibinfo {author} {\bibfnamefont {V.~N.}\ \bibnamefont
  {Kotov}}, \bibinfo {author} {\bibfnamefont {B.}~\bibnamefont {Uchoa}},
  \bibinfo {author} {\bibfnamefont {V.~M.}\ \bibnamefont {Pereira}}, \bibinfo
  {author} {\bibfnamefont {A.~H.}\ \bibnamefont {Castro-Neto}}, \ and\ \bibinfo
  {author} {\bibfnamefont {F.}~\bibnamefont {Guinea}},\ }\href@noop {} {}
  (\bibinfo {year} {2010}),\ \Eprint {http://arxiv.org/abs/1012.3484}
  {arXiv:1012.3484} \BibitemShut {NoStop}%
\bibitem [{\citenamefont {Meng}\ \emph {et~al.}(2010)\citenamefont {Meng},
  \citenamefont {Lang}, \citenamefont {Wessel}, \citenamefont {Assaad},\ and\
  \citenamefont {Muramatsu}}]{meng_spin_liquid_2010}%
  \BibitemOpen
  \bibfield  {author} {\bibinfo {author} {\bibfnamefont {Z.~Y.}\ \bibnamefont
  {Meng}}, \bibinfo {author} {\bibfnamefont {T.~C.}\ \bibnamefont {Lang}},
  \bibinfo {author} {\bibfnamefont {S.}~\bibnamefont {Wessel}}, \bibinfo
  {author} {\bibfnamefont {F.~F.}\ \bibnamefont {Assaad}}, \ and\ \bibinfo
  {author} {\bibfnamefont {A.}~\bibnamefont {Muramatsu}},\ }\href@noop {}
  {\bibfield  {journal} {\bibinfo  {journal} {Nature}\ }\textbf {\bibinfo
  {volume} {464}},\ \bibinfo {pages} {847} (\bibinfo {year}
  {2010})}\BibitemShut {NoStop}%
\bibitem [{\citenamefont {Esquinazi}\ \emph {et~al.}(2003)\citenamefont
  {Esquinazi}, \citenamefont {Spemann}, \citenamefont {H\"ohne}, \citenamefont
  {Setzer}, \citenamefont {Han},\ and\ \citenamefont
  {Butz}}]{esquinazi_proton_irradiaion_graphite_magnetism_2003}%
  \BibitemOpen
  \bibfield  {author} {\bibinfo {author} {\bibfnamefont {P.}~\bibnamefont
  {Esquinazi}}, \bibinfo {author} {\bibfnamefont {D.}~\bibnamefont {Spemann}},
  \bibinfo {author} {\bibfnamefont {R.}~\bibnamefont {H\"ohne}}, \bibinfo
  {author} {\bibfnamefont {A.}~\bibnamefont {Setzer}}, \bibinfo {author}
  {\bibfnamefont {K.-H.}\ \bibnamefont {Han}}, \ and\ \bibinfo {author}
  {\bibfnamefont {T.}~\bibnamefont {Butz}},\ }\href@noop {} {\bibfield
  {journal} {\bibinfo  {journal} {Phys. Rev. Lett.}\ }\textbf {\bibinfo
  {volume} {91}},\ \bibinfo {pages} {227201} (\bibinfo {year}
  {2003})}\BibitemShut {NoStop}%
\bibitem [{\citenamefont {Son}\ \emph {et~al.}(2006)\citenamefont {Son},
  \citenamefont {Cohen},\ and\ \citenamefont
  {Louie}}]{son_edge_magnetism_2006}%
  \BibitemOpen
  \bibfield  {author} {\bibinfo {author} {\bibfnamefont {Y.-W.}\ \bibnamefont
  {Son}}, \bibinfo {author} {\bibfnamefont {M.~L.}\ \bibnamefont {Cohen}}, \
  and\ \bibinfo {author} {\bibfnamefont {S.~G.}\ \bibnamefont {Louie}},\
  }\href@noop {} {\bibfield  {journal} {\bibinfo  {journal} {Phys. Rev. Lett.}\
  }\textbf {\bibinfo {volume} {97}},\ \bibinfo {pages} {216803} (\bibinfo
  {year} {2006})}\BibitemShut {NoStop}%
\bibitem [{\citenamefont {Kim}\ \emph {et~al.}(2003)\citenamefont {Kim},
  \citenamefont {Choi}, \citenamefont {Chang},\ and\ \citenamefont
  {Tom\'anek}}]{kim_cnt_magnetism_2003}%
  \BibitemOpen
  \bibfield  {author} {\bibinfo {author} {\bibfnamefont {Y.-H.}\ \bibnamefont
  {Kim}}, \bibinfo {author} {\bibfnamefont {J.}~\bibnamefont {Choi}}, \bibinfo
  {author} {\bibfnamefont {K.~J.}\ \bibnamefont {Chang}}, \ and\ \bibinfo
  {author} {\bibfnamefont {D.}~\bibnamefont {Tom\'anek}},\ }\href@noop {}
  {\bibfield  {journal} {\bibinfo  {journal} {Phys. Rev. B}\ }\textbf {\bibinfo
  {volume} {68}},\ \bibinfo {pages} {125420} (\bibinfo {year}
  {2003})}\BibitemShut {NoStop}%
\bibitem [{\citenamefont {Gmitra}\ \emph {et~al.}(2009)\citenamefont {Gmitra},
  \citenamefont {Konschuh}, \citenamefont {Ertler}, \citenamefont
  {Ambrosch-Draxl},\ and\ \citenamefont {Fabian}}]{gmitra_soi_2009}%
  \BibitemOpen
  \bibfield  {author} {\bibinfo {author} {\bibfnamefont {M.}~\bibnamefont
  {Gmitra}}, \bibinfo {author} {\bibfnamefont {S.}~\bibnamefont {Konschuh}},
  \bibinfo {author} {\bibfnamefont {C.}~\bibnamefont {Ertler}}, \bibinfo
  {author} {\bibfnamefont {C.}~\bibnamefont {Ambrosch-Draxl}}, \ and\ \bibinfo
  {author} {\bibfnamefont {J.}~\bibnamefont {Fabian}},\ }\href@noop {}
  {\bibfield  {journal} {\bibinfo  {journal} {Phys. Rev. B}\ }\textbf {\bibinfo
  {volume} {80}},\ \bibinfo {pages} {235431} (\bibinfo {year}
  {2009})}\BibitemShut {NoStop}%
\bibitem [{\citenamefont {Wehling}\ \emph {et~al.}(2011)\citenamefont
  {Wehling}, \citenamefont {Sasioglu}, \citenamefont {Friedrich}, \citenamefont
  {Lichtenstein}, \citenamefont {Katsnelson},\ and\ \citenamefont
  {Bl\"ugel}}]{wehling_strength_U_2011}%
  \BibitemOpen
  \bibfield  {author} {\bibinfo {author} {\bibfnamefont {T.~O.}\ \bibnamefont
  {Wehling}}, \bibinfo {author} {\bibfnamefont {E.}~\bibnamefont {Sasioglu}},
  \bibinfo {author} {\bibfnamefont {C.}~\bibnamefont {Friedrich}}, \bibinfo
  {author} {\bibfnamefont {A.~I.}\ \bibnamefont {Lichtenstein}}, \bibinfo
  {author} {\bibfnamefont {M.~I.}\ \bibnamefont {Katsnelson}}, \ and\ \bibinfo
  {author} {\bibfnamefont {S.}~\bibnamefont {Bl\"ugel}},\ }\href@noop {} {}
  (\bibinfo {year} {2011}),\ \Eprint {http://arxiv.org/abs/1101.4007}
  {arXiv:1101.4007} \BibitemShut {NoStop}%
\bibitem [{\citenamefont {Schmidt}\ and\ \citenamefont
  {Loss}(2010{\natexlab{a}})}]{schmidt_gg_soi_2010}%
  \BibitemOpen
  \bibfield  {author} {\bibinfo {author} {\bibfnamefont {M.~J.}\ \bibnamefont
  {Schmidt}}\ and\ \bibinfo {author} {\bibfnamefont {D.}~\bibnamefont {Loss}},\
  }\href@noop {} {\bibfield  {journal} {\bibinfo  {journal} {Phys. Rev. B}\
  }\textbf {\bibinfo {volume} {81}},\ \bibinfo {pages} {165439} (\bibinfo
  {year} {2010}{\natexlab{a}})}\BibitemShut {NoStop}%
\bibitem [{\citenamefont {Ando}(2000)}]{ando_soi_cnt_2000}%
  \BibitemOpen
  \bibfield  {author} {\bibinfo {author} {\bibfnamefont {T.}~\bibnamefont
  {Ando}},\ }\href@noop {} {\bibfield  {journal} {\bibinfo  {journal} {J. Phys.
  Soc. Jpn.}\ }\textbf {\bibinfo {volume} {69}},\ \bibinfo {pages} {1757}
  (\bibinfo {year} {2000})}\BibitemShut {NoStop}%
\bibitem [{\citenamefont {Schmidt}\ and\ \citenamefont
  {Loss}(2010{\natexlab{b}})}]{schmidt_tem_2010}%
  \BibitemOpen
  \bibfield  {author} {\bibinfo {author} {\bibfnamefont {M.~J.}\ \bibnamefont
  {Schmidt}}\ and\ \bibinfo {author} {\bibfnamefont {D.}~\bibnamefont {Loss}},\
  }\href@noop {} {\bibfield  {journal} {\bibinfo  {journal} {Phys. Rev. B}\
  }\textbf {\bibinfo {volume} {82}},\ \bibinfo {pages} {085422} (\bibinfo
  {year} {2010}{\natexlab{b}})}\BibitemShut {NoStop}%
\bibitem [{\citenamefont {Klinovaja}\ \emph
  {et~al.}(2011{\natexlab{a}})\citenamefont {Klinovaja}, \citenamefont
  {Schmidt}, \citenamefont {Braunecker},\ and\ \citenamefont
  {Loss}}]{klinovaja_cnt_long_2011}%
  \BibitemOpen
  \bibfield  {author} {\bibinfo {author} {\bibfnamefont {J.}~\bibnamefont
  {Klinovaja}}, \bibinfo {author} {\bibfnamefont {M.~J.}\ \bibnamefont
  {Schmidt}}, \bibinfo {author} {\bibfnamefont {B.}~\bibnamefont {Braunecker}},
  \ and\ \bibinfo {author} {\bibfnamefont {D.}~\bibnamefont {Loss}},\
  }\href@noop {} {} (\bibinfo {year} {2011}{\natexlab{a}}),\ \Eprint
  {http://arxiv.org/abs/1106.3332} {arXiv:1106.3332} \BibitemShut {NoStop}%
\bibitem [{\citenamefont {Serrano}\ \emph {et~al.}(2000)\citenamefont
  {Serrano}, \citenamefont {Cardona},\ and\ \citenamefont
  {Ruf}}]{serrano_soi_strength_abinitio_2000}%
  \BibitemOpen
  \bibfield  {author} {\bibinfo {author} {\bibfnamefont {J.}~\bibnamefont
  {Serrano}}, \bibinfo {author} {\bibfnamefont {M.}~\bibnamefont {Cardona}}, \
  and\ \bibinfo {author} {\bibfnamefont {J.}~\bibnamefont {Ruf}},\ }\href@noop
  {} {\bibfield  {journal} {\bibinfo  {journal} {Solid State Commun.}\ }\textbf
  {\bibinfo {volume} {113}},\ \bibinfo {pages} {411} (\bibinfo {year}
  {2000})}\BibitemShut {NoStop}%
\bibitem [{\citenamefont {Fujita}\ \emph {et~al.}(1996)\citenamefont {Fujita},
  \citenamefont {Wakabayashi}, \citenamefont {Nakada},\ and\ \citenamefont
  {Kusakabe}}]{fujita_edge_states_1996}%
  \BibitemOpen
  \bibfield  {author} {\bibinfo {author} {\bibfnamefont {M.}~\bibnamefont
  {Fujita}}, \bibinfo {author} {\bibfnamefont {K.}~\bibnamefont {Wakabayashi}},
  \bibinfo {author} {\bibfnamefont {K.}~\bibnamefont {Nakada}}, \ and\ \bibinfo
  {author} {\bibfnamefont {K.}~\bibnamefont {Kusakabe}},\ }\href
  {http://jpsj.ipap.jp/link?JPSJ/65/1920/} {\bibfield  {journal} {\bibinfo
  {journal} {J. Phys. Soc. Jpn.}\ }\textbf {\bibinfo {volume} {65}},\ \bibinfo
  {pages} {1920} (\bibinfo {year} {1996})}\BibitemShut {NoStop}%
\bibitem [{\citenamefont {Luitz}\ \emph {et~al.}(2011)\citenamefont {Luitz},
  \citenamefont {Assaad},\ and\ \citenamefont {Schmidt}}]{luitz_ed_2011}%
  \BibitemOpen
  \bibfield  {author} {\bibinfo {author} {\bibfnamefont {D.~J.}\ \bibnamefont
  {Luitz}}, \bibinfo {author} {\bibfnamefont {F.~F.}\ \bibnamefont {Assaad}}, \
  and\ \bibinfo {author} {\bibfnamefont {M.~J.}\ \bibnamefont {Schmidt}},\
  }\href@noop {} {\bibfield  {journal} {\bibinfo  {journal} {Phys. Rev. B}\
  }\textbf {\bibinfo {volume} {83}},\ \bibinfo {pages} {195432} (\bibinfo
  {year} {2011})}\BibitemShut {NoStop}%
\bibitem [{Note1()}]{Note1}%
  \BibitemOpen
  \bibinfo {note} {The Dirac points $\protect \frac {2\pi }3,\protect \frac
  {4\pi }3$ are excluded explicitly in this work by considering CNT
  circumferences $N_c$ which are not multiples of 3.}\BibitemShut {Stop}%
\bibitem [{\citenamefont {Klinovaja}\ \emph
  {et~al.}(2011{\natexlab{b}})\citenamefont {Klinovaja}, \citenamefont
  {Schmidt}, \citenamefont {Braunecker},\ and\ \citenamefont
  {Loss}}]{klinovaja_helical_modes_2011}%
  \BibitemOpen
  \bibfield  {author} {\bibinfo {author} {\bibfnamefont {J.}~\bibnamefont
  {Klinovaja}}, \bibinfo {author} {\bibfnamefont {M.~J.}\ \bibnamefont
  {Schmidt}}, \bibinfo {author} {\bibfnamefont {B.}~\bibnamefont {Braunecker}},
  \ and\ \bibinfo {author} {\bibfnamefont {D.}~\bibnamefont {Loss}},\
  }\href@noop {} {\bibfield  {journal} {\bibinfo  {journal} {Phys. Rev. Lett.}\
  }\textbf {\bibinfo {volume} {106}},\ \bibinfo {pages} {156809} (\bibinfo
  {year} {2011}{\natexlab{b}})}\BibitemShut {NoStop}%
\end{thebibliography}%

\end{document}